\documentclass[12pt]{iopart}

\usepackage{iopams}  

\usepackage{color}
\usepackage{tikz}
\usetikzlibrary{matrix}

\begin{document}

\title[Spatially non-uniform phases of matter with no long-range order]{A measure theoretical approach to non-uniform phases of matter with no long-range spatial order}

\author{Gyula I. T\'oth}

\address{Interdisciplinary Centre for Mathematical Modelling, Department of Mathematical Sciences, Loughborough University, Epinal Way, LE11 3TU, Loughborough, United Kingdom}
\ead{g.i.toth@lboro.ac.uk}
\vspace{10pt}
\begin{indented}
\item[]July 2020
\end{indented}

\begin{abstract}
In this paper, the development of a mathematical method is presented to explore spatially non-uniform phases with no long-range order in mathematical models of first order phase transitions. We use essential results regarding the concentration of measure phenomenon to re-formulate partial differential equations for probability measures, which extends the concept of analytical solutions to random fields. A stochastic solution of an equation is such a non-singular probability measure, according to which the random variable is almost surely a solution to the equation. The general concept is applied for continuum theories, where the concept of symmetry breaking is extended to probability measures. The concept is practicable and predictive for non-local continuum mean-field theories of first order phase transitions. The results suggest that symmetry breaking must be present in stochastic stationary points of the energy on the level of the probability measure. This is in agreement with the observation that amorphous solid structures preserve local ordering. 
\end{abstract}

%
\vspace{2pc}

%
%
%
%

\section{Introduction}

Prior to the discovery of topological phase transitions, symmetry breaking was the major concept of classifying transition processes between phases of matter. From the mathematical point of view, describing symmetry breaking transitions in physical systems often manifests in finding bifurcation points in the parameter space of partial differential equations (PDE) in the corresponding mathematical model. The general experience is that first order phase transitions are accompanied by symmetry breaking, and in many systems multiple symmetry-breaking phases exist. The one with the minimal energy gives the ground state, while others are metastable phases (i.e., local energy minima). Classifying physical phases in the spirit of symmetry breaking is robust and convenient as long as we only consider "well-behaved" solutions, such as homogeneous, crystal lattice symmetric, etc. Nevertheless, experiments \cite{amorphousice,Chen7685,cahn_bendersky_2003,BENDERSKY2012S171,PhysRevLett.111.015502} and computer simulations \cite{PhysRevLett.107.175702} have provided firm evidence that spatially non-uniform metastable phases with no long-range order (called amorphous solid phase henceforth) can form in first-order phase transitions. In continuum pattern formation models such a phase consists of (infinitely) many different ``random'' configurations represented by smooth random fields (see Figure 1). The energy density distribution of the configurations show a sharpening distribution with size with converging expectation value \cite{Shaho}. Since these solutions are neither disordered (homogeneous) nor ordered (crystalline) in the usual sense, and it is hard to decide what symmetry of is broken here. If one said that all the symmetries are broken, then the question would arise: Why does the crystalline phase have the lowest energy? The consistent inclusion of these solutions in the framework of symmetry breaking is therefore non-trivial. To further complicate the situation, the question of the existence of non-crystalline solid phases forming in first order phase transitions is inseparable from one of the most controversial topic in physical research, the glass transition theory.

\begin{figure}
\includegraphics[width=0.5\linewidth]{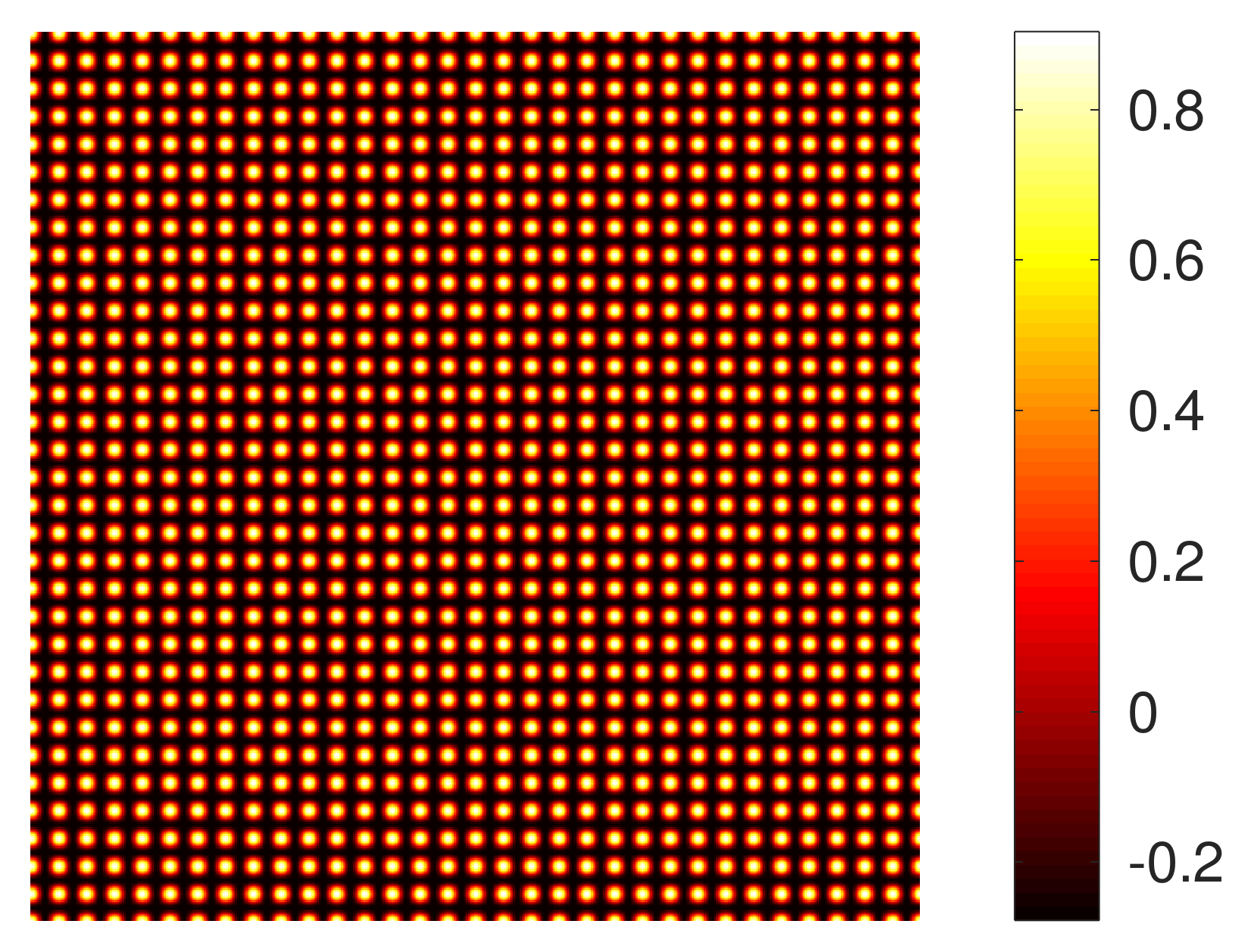}
\includegraphics[width=0.5\linewidth]{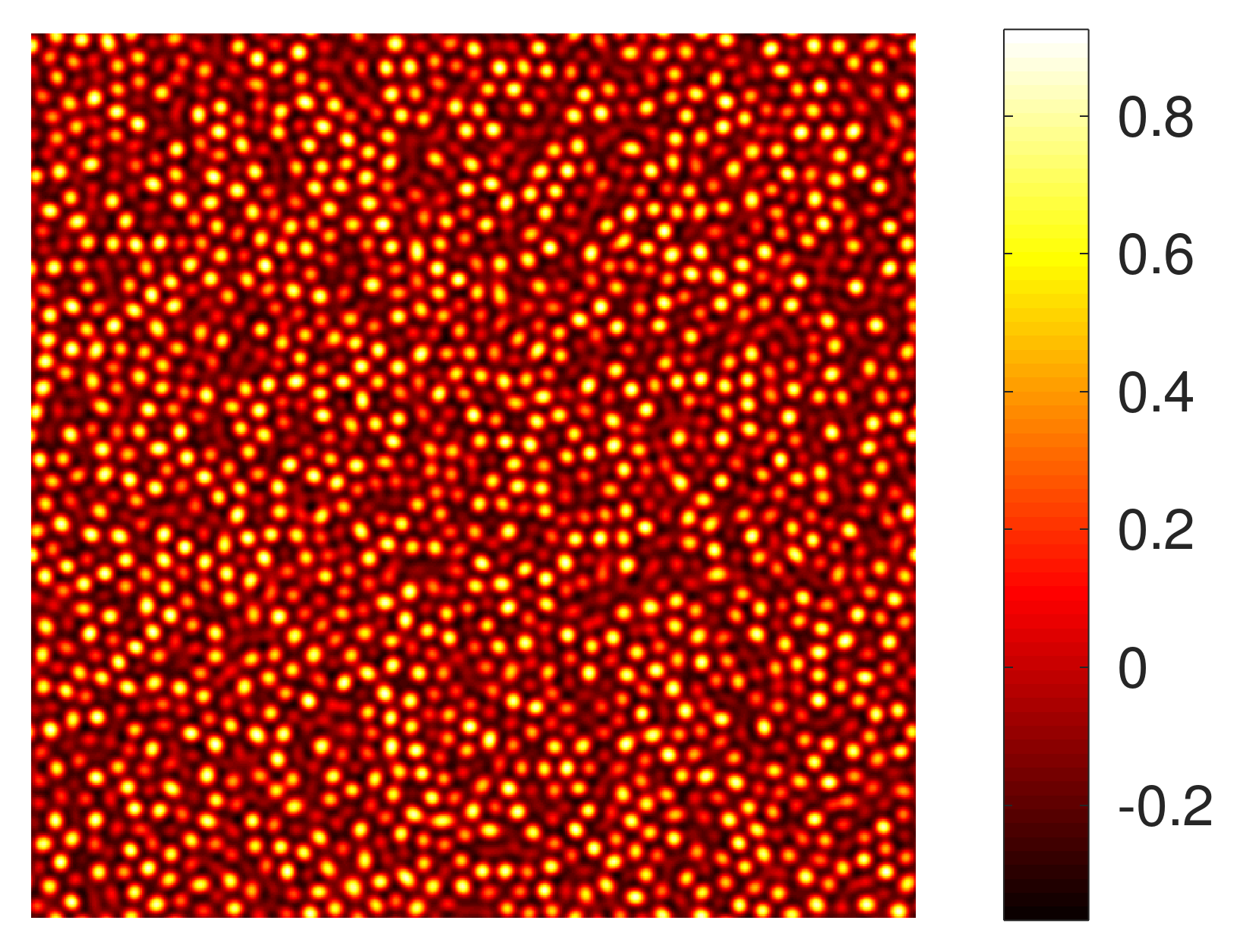}
\caption{Two-dimensional cross section of three-dimensional symmetry breaking numerical solutions of the equation $[(1+\nabla^2)^2-\epsilon]\phi+(3\,\bar{\varphi})\phi^2+\phi^3=\mu$ for zero spatial average at $\bar{\varphi}=-0.25$ and $\epsilon=0.01875$. Besides the continuous symmetry preserving solution $\phi_0(\mathbf{r})=0$, symmetry breaking solutions with body-centred cubic (bcc) symmetry (on the left) and with no long-range order (on the right) also emerge. The solutions represent minima of the functional $F = \int d^3x \,\left\{\frac{1}{2}\phi[(1+\nabla^2)^2-\epsilon]\phi + \bar{\varphi}\,\phi^3 + \frac{1}{4}\phi^4 \right\}$ for the condition $\int d^3x\,\phi = 0$. While the bcc symmetric solution represents the absolute minimum of the energy functional, the solution displaying no long range order has a slightly higher energy and therefore it corresponds to a local energy minimum.}
\end{figure}

The main obstacle towards a theoretical understanding of the fundamental nature of non-uniform structures with no long range order (such as amorphous materials or structural glasses) and transition phenomena involving them is that there exist no mathematical tools even to confirm or disprove their existence in mathematical models \cite{Melbourne1998}. In the lack of such mathematical tool, the standard method of describing disordered phases analytically is to reduce the complexity of the model by parametrising the domain (usually a function space) by using a probability measure characterised by a few scalar parameters \cite{doi:10.1146/annurev.physchem.58.032806.104653}. The expectation value of the energy density then depends on only these parameters, and therefore its stationary points can be easily found (either analytically or numerically). To confirm the assumptions and the approximations made in reducing the complexity of the original energy density surface, the results are usually compared to the results of numerical simulations providing individual energy minima of the original model. The concept of reducing the complexity of the model on the level of the free energy functional is summarised in Figure 2(a). The main issue with this approach is that the stationary points of the reduced model are not expected to coincide with the stationary points of the original energy density. In fact, the original energy extremum equation is expected to be satisfied only approximately even in the expectation value sense, simply because expectation value and extremum calculation are not commutable. Observations indicate that the deviation between the reduced and the original model may become significant in the physically interesting regime of the parameter space \cite{doi:10.1002/9781118202470.ch2}. 

\begin{figure}
\centering
\begin{tikzpicture}
  \matrix (m) [matrix of math nodes,column sep=8.5em,minimum width=2em]
  {
     \textrm{(a)}\quad\enskip\omega[x]\enskip & \enskip f(\vec{\alpha}) := \langle \omega[x] \rangle\enskip & \enskip \langle\delta\omega[X]\rangle \approx 0 \enskip \\};
  \path[-stealth]
    (m-1-1) edge node [above] {$d\mu(\vec{\alpha},x)$} (m-1-2)
    (m-1-2) edge node [above] {$\partial f / \partial \alpha_i =0$} (m-1-3);
\end{tikzpicture}
\begin{tikzpicture}
  \matrix (m) [matrix of math nodes,column sep=7em,minimum width=2em]
  {
     \textrm{(b)}\quad\enskip\delta\omega[x]=0 \enskip & \enskip \mathbb{P}\left( \delta\omega[X]=0 \right)=1 \enskip & \enskip \delta\omega[X] = 0 \enskip \\};
  \path[-stealth]
    (m-1-1) edge node [above] {$d\mu(C,x)$} (m-1-2)
    (m-1-2) edge node [above] {$\mathcal{D}(C)=0$} (m-1-3);
\end{tikzpicture}
\caption{Schematic representation of the (a) traditional and (b) the proposed method of describing stochastic extrema of the energy density $\omega[x]$ of the system. In route (a), the energy density is simplified by calculating its expectation value for a probability measure characterised by a few scalar parameters $\vec{\alpha}$. Consequently, the stationary points of $f(\vec{\alpha})$ do not necessarily coincide with the stationary points of $\omega[x]$, and therefore the energy extremum equation is satisfied only approximately even in the expectation value sense. In contrast, route (b) start from the energy extremum equation, and requires that any random state (represented by $X$) generated according to a parametric measure $d\mu(C,x)$ must be the solution of the equation. If such parameters $C$ exist (where $C$ is a solution of the equation $\mathcal{D}(C)=0$ emerging from the condition), $d\mu(C,x)$ provides an infinite set of exact solutions to the energy extremum equation. The equation $\mathcal{D}(C)=0$ can be a differential equation for a the spatial correlation function $C(\mathbf{r})$ of a smooth random field, for instance.}
\end{figure}

To overcome the problem emerging from the artificial restriction of the domain, here we present a mathematical method, which extends the domain of possible analytical solutions to probability measures, thus providing a tool to study stochastic stationary points of the free energy density. The central idea of the concept is the following. Since physical phases represent stationary points of the free energy density of the system, the configurations forming the amorphous solid phase must be analytical solutions to the energy extremum equation (providing the stationary points of the relevant energy density of the system). Finding an infinite set of ``random'' solutions to an equation is equivalent to finding a non-singular probability measure $d\mu(x)$ on the domain $S$ so that the probability that the random variable $X \in S$ (usually a function or a vector) represents a solution to the equation is unity. If such measure exists, the free energy density is said to have a stochastic stationary point, which represents an amorphous solid phase. The concept is summarised in Figure 2(b). The rest of the paper is structured as follows. In Section 2 we present the general mathematical framework of extending the concept of analytical solution to probability measures. To demonstrate the legitimacy of the idea, an analogy is drawn with the well-known result in statistical physics that the statistical physical ensembles converge in the thermodynamic limit. In Section 3 we apply the general concept to find periodic random field solutions to time-independent partial differential equations. Since we search for symmetry breaking solutions, the consequences of the symmetry properties of differential equations on the solution is extended to probability measures. In Section 4 we investigate whether there exist symmetry-preserving stochastic extrema in non-local mean-field theories with quartic non-linearity. We show that -under certain restrictions - the measure equation is a partial differential equation for the correlation function of a coloured Gaussian random field. For the Swift-Hohenberg operator the equation has an asymptotically vanishing solution in the orders of a small parameter, thus suggesting that the continuous symmetries of the energy functional also must break in stochastic stationary points. Although finding a symmetry breaking stochastic solution is beyond the scope of the present work, we demonstrate that the free energy density and the stability properties of the approximating symmetry-preserving stochastic solution are calculable, thus demonstrating the practicability of the developed mathematical method. The results are summarised in Section 5.

\section{The mathematical framework}

\subsection{Probability Null Measure of operators}

Let $x \in S$ represent the physical state of the system, where $S$ is the set of the possible states. Let $\hat{O}: S \to S$ an operator on the set, and let $\hat{O}[x]=\underline{0}$ the energy extremum equation, where $\underline{0}$ is the additive identity in $S$. Let $\epsilon$ be the null set of $\hat{O}$: $\epsilon:=\{x \in S\,;\enskip \hat{O}[x]=\underline{0} \}$, i.e., $\epsilon$ is the set of solutions to the equation $\hat{O}[x]=\underline{0}$. Assuming that $(S,\Sigma,\mu)$ measure space exists, where $\Sigma$ is a $\sigma$-algebra of $S$ and $\mu$ is a probability measure, $\mu$ is called a \textit{probability null measure} of $\hat{O}$ if and only if $\mu(\epsilon)=1 $ (and therefore $\mu(S \setminus \epsilon)=0$). Using the measure theoretical interpretation of probability, and denoting probability by $\mathbb{P}$, $\mu(\epsilon)=1$ is equivalent to $\mathbb{P}(X \in \epsilon)=1$, meaning that the probability that the random variable $X \in S$ is a solution to the equation $\hat{O}[x]=\underline{0}$ is unity:
\begin{equation}
\label{central}
\mathbb{P}(\hat{O}[X]=\underline{0})=1 \enskip ,
\end{equation}
which emerges from the equivalence of $x \in \epsilon$ and $\hat{O}[x]=\underline{0}$. Consequently, Eq. (\ref{central}) expresses that the random variable chosen from $S$ according to $\mu$ is \textit{almost surely} a solution to the equation $\hat{O}[x]=\underline{0}$. Furthermore, if $\mu$ is non-singular in at least one dimension of the representation of the elements of $S$, $\mu$ is called a \textit{stochastic} solution to the equation $\hat{O}[x]=\underline{0}$, since in this case the generation of the random variable $X$ includes randomness. Here we note that if $\mu$ was concentrated on known solutions to the equation $\hat{O}[x]=\underline{0}$ obtained by traditional equation solving techniques, stochastic solutions would be completely eliminated. These are the trivial solutions of Eq. (\ref{central}) for $\mu$. In contrast, our question here is whether there exists such a non-singular $\mu$ to a given $\hat{O}$ on a given $S$, which generates an infinite set of random solutions (such as a spatially correlated random field in case of a regular partial differential equation) to the equation $\hat{O}[x]=\underline{0}$ . If yes, these sets represent phases of matter with no long range spatial order. 

To apply Eq. (\ref{central}), a practical form of this abstract equation must be given. Let now $||x||^2_2$ be the square of the 2-norm (up to a constant factor) on a representation of $x \in S$. Since $||x||_2^2\geq 0$ and $||x||_2^2 = 0$ if and only if $x = \underline{0}$, requiring
\begin{equation}
\label{practical}
\langle ||\hat{O}[X]||_2^2 \rangle =0
\end{equation} 
(where $\langle.\rangle$ stands for the expectation value) implies Eq. (\ref{central}) for the following reason. $\langle ||\hat{O}[X]||_2^2\rangle =\int_S d\mu(x) \{||\hat{O}[x]||_2^2\}=0$ means that $||\hat{O}[x]||_2^2=0$ ``almost everywhere'', or, in other words, $\mu(\chi)=0$, where $\chi:=\{x \in S\,;\enskip ||\hat{O}[x]||_2^2\neq 0\}$. The equivalence $||y||_2^2=0 \,\Leftrightarrow\,y=\underline{0}$ (or $||y||_2^2 \neq 0 \,\Leftrightarrow\,y \neq \underline{0}$) yields $\chi \equiv S \setminus \epsilon$, and therefore $\mu(\chi)=\mu(S \setminus \epsilon)=0$, thus implying $\mu(\epsilon)=1$, which is equivalent to Eq. (\ref{central}). We note that Eq. (\ref{practical}) is more practical than Eq. (\ref{central}), since it contains the moments $\langle X_i \rangle$, $\langle X_i X_j \rangle$, $
\langle X_i X_j X_k \rangle$, where $X_l$ stands for an element of $X$ in a particular representation.\\

\noindent\textit{Comment: The measure theoretical meaning of the term ``almost surely'' is $\mu(S \setminus \epsilon)=0$, which is equivalent to $\mathbb{P}(X \in S \setminus \epsilon)=0$, i.e., the probability that $X$ is not a solution to the equation $\hat{O}[x]=\underline{0}$ is $0$, and therefore it is an impossible event. We note here that the probability density function (if it exists) is not necessarily zero for impossible events. As a trivial example, one may consider the 
 standard normal distribution on $\mathbb{R}$. Even if $p(x)\neq 0$ for $x \in \mathbb{Z}$, the probability that the random variable $X$ takes an integer value is zero. This is obvious from the following discretisation of the continuous distribution. Let the probability mass function be $P_n(x_k):=(2^n \sqrt{2\,\pi})^{-1}\exp(-x_k^2/2)$, where $x_k = k/2^n$, $k \in \mathbb{Z}$ and $n \in \mathbb{N} \cup \{0\}$. The probability that $X_n$ is integer reads: $\mathbb{P}(X_n \in \mathbb{Z}) = \left[\sum_{k=-\infty}^\infty P_n(k)\right]/\left[ \sum_{k=-\infty}^\infty P_n(k/2^n) \right] = 1/2^n$. This indicates $\lim_{n \to \infty} \mathbb{P}(X_n \in \mathbb{Z})=0$, even though $p(x)=\lim_{n \to \infty} 2^n P_n(x) = (\sqrt{2\,\pi})^{-1}\exp(-x^2/2)$ is non-zero for $x \in \mathbb{Z}$. The result emerges from the fact that the density of $\mathbb{Z}$ in $\mathbb{R}$ is $0$.}

\subsection{Equivalence of statistical physical ensembles in the thermodynamic limit}     

Before applying Eq. (\ref{practical}) to specific models, first we need to address the mathematical adequacy of Eq. (\ref{central}). For a non-singular measure for which Eq. (\ref{central}) holds, the operator $\hat{O}$  transforms ``true'' random states into $\underline{0}$. This phenomenon is known in mathematics as the limit case of the Concentration of Measure Phenomenon (CMP) \cite{MilmanLevy,Milman2010TheCP,Talagrand1995,ledoux2001concentration}, the generalisation of the Central Limit Theorem. Roughly speaking, the CMP states  that a Lipshitz-continuous function of a large number of independent random variables is nearly constant if the function is balanced, i.e., does not depend ``too much'' on any of the variables. The mathematical formulation of the concept is done via ``concentration inequalities''. Without loss of generality, the essence of these inequalities can be summarised as: $\mathbb{P}\left(|f(\mathbf{X})-\bar{f}|\geq\lambda\right)\leq C\,\exp(-\alpha_N \lambda^p)$ (where $C>0$ and $p>0$), i.e., the probability that the random variable $Y:=f(\mathbf{X})$ (where $\mathbf{X}$ denotes $N$ random variables) is outside of the $\lambda$ vicinity of a suitable median (usually the expectation value) $\bar{f}$ decays exponentially with $\lambda$, while $\alpha_N$ is a positive and strictly increasing function of $N$. If $\lim_{N \to \infty} \alpha_N \to \infty$, then $\lim_{N \to \infty} \mathbb{P}(|f(\mathbf{X})-\bar{f}|>\lambda)=0$ for any $\lambda>0$, and therefore $\mathbb{P}\left( f(\mathbf{X})=\bar{f} \right)=1$, i.e., the random variable $Y=f(\mathbf{X})$ becomes ``almost surely'' constant in this limit. (Since $\mathbf{X}$ is a random vector in the infinite dimensional Euclidean space, which is measurable in the probability sense, the $N \to \infty$ limit can be taken.) This result means that any balanced and Lipshitz continuous function of infinitely many independent random variables is constant. 

Based on the above argumentation, it is reasonable to ask whether Eq. (\ref{central}) holds for non-singular probability measures. We note, however, that our problem is the inverse problem, i.e., a measure is to be found to a given function of the random variables. Due to the presence of physical interactions the function is usually not a Lipshitz function of the variables. Consequently, there is no guarantee that such a measure exist or that it is a product measure. Nevertheless, an example where the solution exists is the "equivalence of statistical physical ensembles in the thermodynamic limit" \cite{gibbs1902elementary,KhinchinGamow}. In particular, the relative amplitude of energy fluctuations vanish for an infinitely large system (with constant density), i.e., $\lim_{N \to \infty}\langle [E(\mathbf{X})/\bar{E}-1]^2 \rangle = 0$ for the Gibbs measure, the energy density $e:=E/V$ is unique: $\mathbb{P}\left(e(\mathbf{X})=\bar{e}\right)=1$, and therefore the canonical ensemble converges to the microcanonical one in the thermodynamic limit. Summarizing, it seems that Eq. (\ref{central}) may apply to non-singular measures if the representations of $S$ are (at least) countable infinitely many dimensional.\\

\textit{Comment: In the above example the function is balanced (the Hamiltonian is invariant under the permutation of particles), however, only locally Lipshitz continuous in the kinetic energy, and might not be Lipshitz continuous in the pair potential (singularities may occur at equal particle positions). Nevertheless, there exists a measure to the Hamiltonian, which is fully concentrated for $N \to \infty$ (for $N/V$ constant), even though the Gibbs measure is not a product measure in the potential energy in general (i.e., the positions are not independent random variables, which directly emerges from the presence of particle interactions). It is important to see that the lack of Lipschitz continuity does not necessarily indicate the non-existence of a measure for which the function is concentrated. In the present example, there exists a non-product probability measure to a non-Lipschitz continuous function, for which the function is concentrated.}

\section{Application: amorphous solid phases in continuum theories}

\subsection{Fourier representation of PDE's}

Let the equation $\hat{O}[\phi]=0$ be a time-independent Partial Differential Equation (e.g. an energy extremum equation), where $\phi(\mathbf{r}) \in \mathbb{R}$ is a scalar field in $d$ spatial dimensions. To investigate the possible stochastic solutions of the equation, in the spirit of Eq. (\ref{practical}) first we need to define a measurable set of functions $S$ with a norm. Let now $S$ be the set of those zero-average box periodic functions in $d$ spatial dimensions, for which the Fourier coefficients exist. The wave numbers are discrete and read: $\mathbf{k}=(2\,\pi/L)\,\mathbf{n}\neq \mathbf{0}$, where $L$ is the period and $\mathbf{n} \in \mathbb{Z}^d$, and any element of $S$ can be represented by the set of its Fourier coefficients. Since $\mathbb{Z}^d \setminus \{\mathbf{0}\}$ is a countable set, the Fourier coefficients can be listed: $\Phi:=\{\phi_{\mathbf{k}_1},\phi_{\mathbf{k}_2},\dots\}$, where $\mathbf{k}_i \neq \mathbf{0}$. In addition, $\phi_{\mathbf{k}_j}=\phi^*_{\mathbf{k}_i}$ for $\mathbf{k}_j+\mathbf{k}_i=\mathbf{0}$ [Hermitian symmetry due to $\phi(\mathbf{r}) \in \mathbb{R}$], and therefore the list can be re-arranged as $\{\phi_{\mathbf{k}_1},\phi_{-\mathbf{k}_1},\phi_{\mathbf{k}_2},\phi_{-\mathbf{k}_2},\dots\}=\{\phi_{\mathbf{k}_1},\phi^*_{\mathbf{k}_1},\phi_{\mathbf{k}_2},\phi^*_{\mathbf{k}_2},\dots\}$, which contains redundant information. Keeping only the real part ($\alpha_{\mathbf{k}_i}$) of $\phi_{\mathbf{k}_i}$ and minus the imaginary part ($\beta_{\mathbf{k}_i}$) of $\phi^*_{\mathbf{k}_i}$ results in $\zeta:=\{ \alpha_{\mathbf{k}_1},\beta_{\mathbf{k}_1},\alpha_{\mathbf{k}_2},\beta_{\mathbf{k}_2}, \dots \}$, where all wave numbers are substantially different (defined by $\mathbf{k}_i \neq \pm \mathbf{k}_j$ for $i \neq j$). Since $\zeta$ is a representation of the countable infinite dimensional Euclidean space, which is measurable in the probability sense, $S$ is measurable in the probability sense. In addition, since the representation of $S$ is infinite dimensional, non-singular measures may satisfy Eq. (\ref{central}). Requiring $\hat{O}:S \to S$, the Fourier representation of the operator is a complex vector-vector function $\Psi:=\mathbf{F}(\Phi)$, where $\Psi=\{\psi_{\mathbf{k}_1},\psi_{\mathbf{k}_2},\psi_{\mathbf{k}_3},\dots\}$ is the ordered list of the the Fourier coefficients of $\psi(\mathbf{r}):=\hat{O}[\phi(\mathbf{r})]$, which then defines $\mathbf{F}$. The Fourier representation of the original PDE reads:
\begin{equation}
\mathbf{F}(\Phi)=\mathbf{0} \enskip .
\end{equation}
A suitable norm on $S$ can be defined as $||\phi||_2^2:=\frac{1}{2\,L^d}\int_{\Omega_L} d^dx \{\phi^2(\mathbf{r})\}$ (where $\Omega_L$ is an arbitrary period in $\mathbb{R}^d$), which is equivalent to $||\Phi||_2^2:=\sum_{i=1}^\infty \phi^*_{\mathbf{k}_i} \phi_{\mathbf{k}_i} = \sum_{i=1}^\infty (\alpha^2_{\mathbf{k}_i} + \beta^2_{\mathbf{k}_i})$ in the Fourier representation. Finally, requiring 
\begin{equation}
\label{psinorm}
\langle F_i^*(\Phi)F_i(\Phi) \rangle =0
\end{equation}
(where $i \in \mathbb{N}$) implies $\langle ||\hat{O}[\phi]||_2^2 \rangle = \langle ||\mathbf{F}(\Phi)||_2^2 \rangle =\sum_{i=1}^\infty \langle F_i^*(\Phi)F_i(\Phi) \rangle = 0$. Consequently, if the probability measure $d\mu(\Phi)$ satisfies Eq. (\ref{psinorm}), it is a solution to Eq. (\ref{practical}) and therefore to Eq. (\ref{central}) too. For the sake of simplicity, in the rest of the paper we use the representation $\Phi:=\{\phi_{\mathbf{k}_1},\phi_{\mathbf{k}_2},\dots\}$, where all wave numbers are substantially different, i.e., $\mathbf{k}_j \neq \pm\mathbf{k}_i$ for $i \neq j$. This is equivalent to eliminating the redundant information emerging from the Hermitian symmetry $\phi_{-\mathbf{k}} \equiv \phi^*_{\mathbf{k}}$. 

\subsection{Symmetry preserving measures}
 
In the framework of the present study we investigate the existence of such solution(s) of Eq. (\ref{psinorm}) which preserves the symmetry properties of $\mathbf{F}(\Phi)=\mathbf{0}$ in the following sense. Let $d\mu(\Phi)$ is a solution to Eq. (\ref{psinorm}). If Eq. (\ref{psinorm}) is formally invariant for the symmetry operation $\Phi \to \hat{T}[\Phi]$, then - analogously to algebraic and differential equations - $d\mu'(\Phi):=d\mu(\hat{T}[\Phi])$ is also a solution of Eq. (\ref{psinorm}). A symmetry preserving solution is defined as $d\mu(\Phi)=d\mu(\hat{T}[\Phi])$, i.e., the solution itself is also invariant for the symmetry operation. If $\hat{O}[\phi]=0$ has continuous translation symmetry in real space, for instance, the invariance of the measure for this symmetry operation is that $\mathbb{P}[\Phi=\phi(\mathbf{r}+\mathbf{d})]$ is constant for any $\mathbf{d} \in \mathbb{R}^d$, i.e., the measure is also invariant for the continuous shift of the so-ordinate system. The same can be applied to rotation and mirror symmetries. We need to mention that requiring these symmetries for $d\mu(\Phi)$ is not necessary: analogously to symmetry breaking regular solutions of the original PDE, symmetry breaking might be present in stochastic solutions on the level of $d\mu(\Phi)$. Nevertheless, in the present study we only demonstrate the practicability of the mathematical concept, and therefore we only consider symmetry preserving stochastic solutions, which is the stochastic analogous of the homogeneous phase.

Let now $\hat{O}$ be isotropic and translation invariant. Since $\hat{O}[\phi]=0$ is invariant for $\mathbf{r} \to \mathbf{r}+\mathbf{d}$ for any $\mathbf{d} \in \mathbb{R}^d$, and the symmetry operation manifests as $\phi_\mathbf{k} \to e^{\imath\,\mathbf{k}\cdot\mathbf{d}}\,\phi_\mathbf{k}$ in the Fourier space, the invariance prescribes
\begin{equation}
\label{translation}
p^{\mathbf{k}_{s_1},\mathbf{k}_{s_2},\dots,\mathbf{k}_{s_n}}_n(\tilde{\phi}_1,\tilde{\phi}_2,\dots,\tilde{\phi}_n) = p^{\mathbf{k}_{s_1},\mathbf{k}_{s_2},\dots,\mathbf{k}_{s_n}}_n(\phi_1,\phi_2,\dots,\phi_n)
\end{equation}
for the $n$-point marginal probability density function (p.d.f.) to an arbitrary set of different wave numbers $\{\mathbf{k}_{s_1},\mathbf{k}_{s_2},\dots,\mathbf{k}_{s_n}\}$, where $\tilde{\phi}_i=e^{\imath\,\mathbf{k}_{s_i}\cdot\mathbf{d}}\phi_i$ with $\mathbf{d} \in \mathbb{R}^d$ also arbitrary. Eq. (\ref{translation}) indicates that the relevant variables of the marginal p.d.f.'s are the rotation invariant amplitude squares $A_{\mathbf{k}_i}^2:=\phi^*_{\mathbf{k}_{i}}\phi_{\mathbf{k}_{i}}$, since $\tilde{A}^2_{\mathbf{k}_i}\equiv A^2_{\mathbf{k}_i}$ for any $i \in \mathbb{N}$ and $\mathbf{d} \in \mathbb{R}^d$. For $p_1^{\mathbf{k}}(\phi)$ it prescribes rotational symmetric bivariate p.d.f. in the real and imaginary components of $\phi$. This means that the amplitude and the phase of $\phi$ are independent, and the phase follows uniform distribution \cite{bryc1995normal}. Furthermore, Maxwell's Theorem states  that the only solution for $p_1^{\mathbf{k}_i}(\phi)$ where the real and imaginary components are independent is the Gaussian distribution. To further restrict the measure, we use the mirror and rotation symmetries of the PDE. With periodic boundary conditions, the PDE is invariant for any rotation that transforms the Fourier space into itself. Combining this with $d$-dimensional mirror symmetry results in that the $n$-point marginal distributions are permutation invariant in arguments for which the wave numbers have equal length. This means that the $p^{\mathbf{k}_i}_1(\phi)=g_{|\mathbf{k}_i|}(A^2)$ (where $A^2=\phi^*\phi$), i.e., the distribution depends only the wave number magnitude.\\

\textit{Comment: The concept of investigating the effect of symmetries of algebraic or differential equations on the structure of the solutions can be extended to Eq. (\ref{psinorm}) as follows. Let $d\mu(\Phi)$ be a solution of Eq. (\ref{psinorm}), i.e., $\int_S d\mu(\phi)\, |F_i(\phi)|^2 =0$, where $|F_i(\phi)|^2:=F_i^*(\phi)F_i(\phi)$. If $|F_i(\phi)|^2$ is invariant for $\phi \to \hat{T}[\phi]$, where the determinant of the Jacobian (denoted by $|J|$) of the transformation $\tilde{\phi}:=\hat{T}[\phi]$ is unity, and $\hat{T}: S \to S$, $d\tilde{\mu}(\phi):=d\mu(\hat{T}[\phi])$ is also a solution of Eq. (\ref{psinorm}), since $$\int_S d\mu(\hat{T}[\phi])\,|F_i(\phi)|^2=\int_{\tilde{S}} d\mu(\tilde{\phi})\,\{ |J|\,|F_i(\hat{T}^{-1}[\tilde{\phi}])|^2 \} = \int_{S} d\mu(\tilde{\phi})\,|F_i(\tilde{\phi})|^2 =0 \enskip .$$ Analogously to the algebraic and differential cases, the result doesn't necessarily indicate that symmetry properties are directly inherited by the solution, only indicates constraints on the structure of the solutions.}

\section{Example: Non-local mean-field theories with quartic non-linearity}

\subsection{An exact result}

In the classical continuum mean-field theories of first order phase transitions with quartic non-linearity, the dimensionless free energy density of the inhomogeneous system is defined as:
\begin{equation}
\label{functional}
f[\varphi] := \frac{1}{L^d}\int_{\Omega_L} d^dx \, \left\{ \frac{1}{2}\varphi\,\hat{\mathcal{L}}[\varphi]+\frac{1}{4}\,\varphi^4 \right\} \enskip ,
\end{equation}
where $\varphi(\mathbf{r}) \in S$ is a scalar field, $S$ the set of box periodic functions in $d$ spatial dimensions, $\Omega_L \in \mathbb{R}^d$ an arbitrary period and $\hat{\mathcal{L}}[\varphi]:=(h * \varphi)(\mathbf{r})$ a convolution with $h(r)$ being an isotropic kernel. Since $\varphi(\mathbf{r})$ often represents a globally conserved physical quantity, we also require $\int_{\Omega_L} d^dx \, \{\varphi-\bar{\varphi} \} \equiv 0$, where $\bar{\varphi}$ is the fixed spatial average of $\varphi(\mathbf{r})$. The simplest examples for the kernel are $h(r)=-(1+\nabla^2)\delta(\mathbf{r})$ (Cahn-Hilliard model) and $h(r)=[(1+\nabla^2)^2-\gamma]\delta(\mathbf{r})$ (conserved Brazowskij / Swift-Hohenberg or Phase-Field Crystal model \cite{PhysRevLett.88.245701,PhysRevE.70.051605,PhysRevB.75.064107}), where $\delta(\mathbf{r})$ is the $d$-dimensional Dirac-delta distribution. The stationary points of $f[\varphi]$ can be determined by solving the Euler-Lagrange equation $\delta f[\varphi] = \Lambda$, where $\delta f[\varphi]$ denotes the first functional derivative of $f[\varphi]$ with respect to $\varphi$, while $\Lambda \in \mathbb{R}$ is a Lagrange multiplier emerging from the global conservation of $\varphi(\mathbf{r})$. Introducing $\phi:=\varphi-\bar{\varphi}$ and re-arranging the equation results in:
\begin{equation}
\label{operator}
(\hat{\mathcal{L}}+3\,\bar{\varphi}^2)\,\phi +(3\,\bar{\varphi})\,\phi^2 + \phi^3 = \nu \enskip ,
\end{equation}
where $\int d^dx\,\phi = 0$ and $\nu \in \mathbb{R}$. Eq. (\ref{operator}) indicates the operator $\hat{O}[\phi]:=(\hat{\mathcal{L}}+3\,\bar{\varphi}^2)\,\phi +(3\,\bar{\varphi})\,\phi^2 + \phi^3 - \nu$ with null set $\epsilon:=\{ \phi \in S \, ; \enskip \hat{O}[\phi]=0 \}$ on $S$. It is trivial to see that $\hat{O}$ is isotropic and translation invariant. Using the Fourier representation of $\phi(\mathbf{r})$ in Eq. (\ref{operator}) yields:
\begin{eqnarray}
\nonumber\psi_\mathbf{q} := && [\mathcal{L}(q)+3\,\bar{\varphi}^2]\,\phi_{\mathbf{q}} + (3\,\bar{\varphi})\sum_{\mathbf{q}_1,\mathbf{q}_2} \phi_{\mathbf{q}_1}\phi_{\mathbf{q}_2} \delta_{\mathbf{q}_1+\mathbf{q}_2-\mathbf{q}} \\ 
\label{psik}&& + \sum_{\mathbf{q}_1,\mathbf{q}_2,\mathbf{q}_3} \phi_{\mathbf{q}_1}\phi_{\mathbf{q}_2}\phi_{\mathbf{q}_3} \delta_{\mathbf{q}_1+\mathbf{q}_2+\mathbf{q}_3-\mathbf{q}} + \nu \,\delta_{\mathbf{q}} = 0 \enskip ,
\end{eqnarray}
where $\mathcal{L}(q)=(2\,\pi)^d h(q)$ with $h(q)$ being the Fourier Transform of $h(r)$, $\delta_{\mathbf{q}}$ stands for the Kronecker symbol giving $1$ for $\mathbf{q}=\mathbf{0}$ and $0$ otherwise. It is easy to see that $\psi_{-\mathbf{q}}=\psi_{\mathbf{q}}^*$ and $\psi_{\mathbf{q}}(\Phi) \equiv 0$ for $\mathbf{q} \neq (2\,\pi/L)\,\mathbf{n}$, where $\mathbf{n} \in \mathbb{Z}^d$, and therefore $\hat{O}: S \to S$ holds. Since continuous translation, discrete rotation and mirror symmetries apply to Eq. (\ref{operator}) for box periodic $\phi(\mathbf{r})$, the results of Section 3.2 can be used. In the simplest symmetry preserving measure 
\begin{enumerate}
\item $\alpha_\mathbf{q}$ and $\beta_\mathbf{q}$ are independent;
\item substantially different wave numbers are independent.
\end{enumerate}
We note that the above conditions are not required on a physical basis, they are motivated solely by mathematical simplicity. Condition (i) indicates that the marginal distributions are Gaussian with variance $\sigma_\mathbf{q}^2 := \langle \phi_{\mathbf{q}}^* \phi_{\mathbf{q}}\rangle = \langle \alpha^2_{\mathbf{q}} \rangle + \langle \beta^2_{\mathbf{q}}\rangle$ and $\langle \alpha^2_{\mathbf{q}} \rangle = \langle \beta^2_{\mathbf{q}}\rangle$, where $\sigma^2_{\mathbf{q}}$ depends only on the wave number magnitude (isotropic), while condition (ii) implies that any $n$-point marginal p.d.f. for substantially different wave numbers is a product measure. Consequently, the measure we consider describes a Gaussian random field, which makes Eq. (\ref{psinorm}) reasonably easily expandable for the only parameter of the measure, the correlation function:
\begin{equation}
\label{isotropic}
C_L(\mathbf{r},\mathbf{r}'):=\langle \phi(\mathbf{r})\phi(\mathbf{r}') \rangle
= \sum_{\mathbf{q},\mathbf{q}'} \langle \phi_{\mathbf{q}} \phi_{\mathbf{q}'} \rangle e^{\imath\,(\mathbf{q}\cdot \mathbf{r}+\mathbf{q}'\cdot \mathbf{r}')} = \sum_{\mathbf{q}}\sigma_\mathbf{q}^2 e^{\imath\,\mathbf{q}\cdot(\mathbf{r}-\mathbf{r}')} \enskip ,
\end{equation}
where $\mathbf{q}$ and $\mathbf{q}'$ run for all possible wave numbers (not only for a set of substantially different ones). We note that $C_L(\mathbf{r},\mathbf{r}')$ is invariant for a continuous shift of the coordinate system, since it only depends on $\mathbf{r}-\mathbf{r}'$, and therefore Condition (ii) is also necessary for a symmetry preserving measure (this also applies to non-Gaussian measures). Furthermore, the isotropy of $\sigma^2_\mathbf{q}$ will result in real-space isotropy in the infinite volume limit: $\lim_{L \to \infty}C_L(\mathbf{r},\mathbf{r}') = C(|\mathbf{r}-\mathbf{r}'|)$. Substituting Eq. (\ref{psik}) into Eq. (\ref{psinorm}) yields:
\begin{eqnarray}
\nonumber && 2 \,[\mathcal{L}(q)+3\,\bar{\varphi}^2] \sum_{\mathbf{q}_1,\mathbf{q}_2,\mathbf{q}_3} \Re \langle \phi_{\mathbf{q}}\phi_{\mathbf{q}_1}\phi_{\mathbf{q}_2}\phi_{\mathbf{q}_3} \rangle \,\delta_{\mathbf{q}_1+\mathbf{q}_2+\mathbf{q}_3+\mathbf{q}}\\
\nonumber && + \sum_{\mathbf{q}_1,\dots,\mathbf{q}_6} \langle \phi_{\mathbf{q}_1}\phi_{\mathbf{q}_2}\phi_{\mathbf{q}_3}\phi_{\mathbf{q}_4}\phi_{\mathbf{q}_5}\phi_{\mathbf{q}_6} \rangle \, \delta_{\mathbf{q}_1+\mathbf{q}_2+\mathbf{q}_3-\mathbf{k}} \, \delta_{\mathbf{q}_4+\mathbf{q}_5+\mathbf{q}_6+\mathbf{q}} \\
\nonumber && + (3\,\bar{\varphi})^2 \sum_{\mathbf{q}_1,\dots,\mathbf{q}_4} \langle \phi_{\mathbf{q}_1}\phi_{\mathbf{q}_2}\phi_{\mathbf{q}_3}\phi_{\mathbf{q}_4} \rangle \, \delta_{\mathbf{q}_1+\mathbf{q}_2-\mathbf{k}} \, \delta_{\mathbf{q}_3+\mathbf{q}_4+\mathbf{k}} \\
\nonumber && + 2\,(3\,\bar{\varphi})[\mathcal{L}+3\,\bar{\varphi}^2]\sum_{\mathbf{q}_1,\mathbf{q}_2} \Re\langle \phi_{\mathbf{k}} \phi_{\mathbf{q}_1}\phi_{\mathbf{q}_2} \rangle \, \delta_{\mathbf{q}_1+\mathbf{q}_2+\mathbf{q}} \\
\nonumber && + 2\,(3\,\bar{\varphi})\sum_{\mathbf{q}_1,\dots,\mathbf{q}_5} \Re\langle \phi_{\mathbf{q}_1} \phi_{\mathbf{q}_2}\phi_{\mathbf{q}_3} \phi_{\mathbf{q}_4} \phi_{\mathbf{q}_5} \rangle \, \delta_{\mathbf{q}_1+\mathbf{q}_2+\mathbf{k}} \,\delta_{\mathbf{q}_3+\mathbf{q}_4+\mathbf{q}_5-\mathbf{q}}\\
\label{fullpsi} && + [\mathcal{L}(q)+3\,\bar{\varphi}^2]^2 \,\langle \phi_{\mathbf{q}}^*\phi_\mathbf{q} \rangle = \omega\,\delta_\mathbf{q} \enskip ,
\end{eqnarray}
where $\Re\langle.\rangle$ stands for the real part of the expectation value, $\omega \in \mathbb{R}$ is responsible for $\langle \psi_\mathbf{0}^* \psi_{\mathbf{0}}\rangle = 0$ for $\phi_\mathbf{0}=0$. To evaluate the moments $\langle \prod_{i=1}^n \phi_{\mathbf{q}_i}\rangle$, we substitute $\phi_{\mathbf{q}}=\alpha_{\mathbf{q}}+\imath\,\beta_{\mathbf{q}}$, then use Isserlis' theorem \cite{10.1093/biomet/12.1-2.134} (also known as Wick's theorem in particle physics). This can be done, since $\alpha_\mathbf{q}$ and $\beta_\mathbf{q}$ follow Gaussian distribution for any $\mathbf{q}$ due to $\phi_{\mathbf{q}'}=\phi^*_{\mathbf{q}}$ for $\mathbf{q}+\mathbf{q}'=\mathbf{0}$. The calculation yields an expression containing only $\langle \alpha_{\mathbf{q}}\alpha_{\mathbf{q}'}\rangle$, $\langle \beta_{\mathbf{q}}\beta_{\mathbf{q}'}\rangle$ and $\langle \alpha_{\mathbf{q}}\beta_{\mathbf{q}'}\rangle$, where $\mathbf{q}$ and $\mathbf{q}'$ can still be arbitrary. Using the $2$-point marginal p.d.f. $p_2(\phi_{\mathbf{k}_i},\phi_{\mathbf{k}_j})=g(A_i,\sigma_i)g(A_j,\sigma_j)$ (where $g(A_i,\sigma_i)$ is the p.d.f. of a zero-mean Gaussian random variable $A_i=\sqrt{\phi^*_{\mathbf{k}_i}\phi_{\mathbf{k}_i}}$ with standard deviation $\sigma_i=\sqrt{\sigma_{\mathbf{k}_i}^2}$ with $\mathbf{k}_i$ and $\mathbf{k}_j$ being substantially different) and $\phi_{-\mathbf{q}}=\phi^*_{\mathbf{q}}$ result in $\langle \alpha_{\mathbf{q}}\beta_{\mathbf{q}'}\rangle=0$ and
\begin{eqnarray}
\langle \alpha_{\mathbf{q}} \alpha_{\mathbf{q}'} \rangle &=& [(\sigma_\mathbf{q}\sigma_{\mathbf{q}'})/2]\left(\delta_{\mathbf{q}-\mathbf{q}'}+\delta_{\mathbf{q}+\mathbf{q}'}\right) \enskip ;\\
\langle \beta_{\mathbf{q}} \beta_{\mathbf{q}'} \rangle &=& [(\sigma_\mathbf{q}\sigma_{\mathbf{q}'})/2]\left(\delta_{\mathbf{q}-\mathbf{q}'}-\delta_{\mathbf{q}+\mathbf{q}'}\right) \enskip ,
\end{eqnarray} 
showing that the real/imaginary parts are correlated/anti-correlated for $\mathbf{q}+\mathbf{q}'=\mathbf{0}$. Using the above correlators results in $\langle \prod_{i=1}^{2m-1} \phi_{\mathbf{q}_i} \rangle = 0$ and $\langle \prod_{i=1}^{2m} \phi_{\mathbf{q}_i} \rangle = \left(\prod_{i=1}^{2m} \sigma_{\mathbf{q}_i} \right) \sum \prod \delta_{\mathbf{q}'+\mathbf{q}''}$, where $m \in \mathbb{N}$, while the notation $\sum\prod$ stands for summing over all distinct ways of partitioning the set $\{\mathbf{q}_1,\mathbf{q}_2,\dots,\mathbf{q}_{2m}\}$ into $m$ pairs $(\mathbf{q}',\mathbf{q}'')$, and each terms is the product of the Kronecker symbol over the pairs. We note that the expectation values exist since the space of the Fourier coefficients is bijective with the infinite dimensional Euclidean space, which is measurable in the probability sense (see the classical Wiener measure, for instance). After lengthy but straightforward algebraic manipulations Eq. (\ref{fullpsi}) reads:  
\begin{eqnarray}
\nonumber \left[ \mathcal{L}(q)+3(\bar{\varphi}^2+c_0) \right]^2\sigma_{\mathbf{q}}^2 + (18\, \bar{\varphi}^2) \sum_{\mathbf{q}_1}\sigma^2_{\mathbf{q}_1}\sigma^2_{\mathbf{q}-\mathbf{q}_1} \\
\label{discrete} + \,6 \sum_{\mathbf{q}_1,\mathbf{q}_2}\sigma^2_{\mathbf{q}_1}\sigma^2_{\mathbf{q}_2}\sigma^2_{\mathbf{q}-(\mathbf{q}_1+\mathbf{q}_2)} = \omega \, \delta_\mathbf{q} \enskip ,
\end{eqnarray}
where $c_0=\sum_{\mathbf{q}}\sigma^2_{\mathbf{q}}$, and $\omega \in \mathbb{R}$ is responsible for the condition $\sigma^2_\mathbf{0}=0$ (emerging from $\phi_\mathbf{0} = 0$). Since $\sigma^2_\mathbf{q} \geq 0$ must hold, the only exact solution of the above equation is $\sigma^2_{\mathbf{q}} = 0$. Since Eq. (\ref{discrete}) is exact for Eq. (\ref{psinorm}), the spatially correlated isotropic Gaussian random field is not a stationary point of Eq. (\ref{functional}). Since the spatially correlated isotropic Gaussian random field preserves all symmetries of the Euler-Lagrange equation on the level of the measure, it is \textit{not} a symmetry breaking stochastic solution in this sense. The importance of this result is inevitable from the physical point of view. The only all-symmetry preserving regular solution to mathematical models of phase transitions is the constant function (representing the uniform physical phase). The first question arising in regards of possible stochastic solutions is the existence of its stochastic analogue. It has been shown that the simplest possible measure preserving all symmetries of the PDE doesn't result in the emergence of a new, all-symmetry preserving stochastic phase. We mention, however, that the spatially correlated Gaussian random field might not be the only all-symmetry preserving measure, and therefore further investigations are needed to confirm / disprove the existence of such a phase. Nevertheless, pur preliminary results suggest that symmetry breaking is necessary for the emergence of a spatially non-uniform phase, either it happens directly in regular solutions or indirectly in probability measures generating the stochastic solutions.      

\subsection{Approximate Gaussian solution in the Swift-Hohenberg model}

\subsubsection{Correlation function.}

The result of the previous section suggest that random field solutions of PDE's contain a symmetry breaking in the probability measure. Though investigating this prediction is well beyond the scope of the present manuscript and is the topic of future research, the first step of the journey is to check whether the idea is practicable at all. For this reason, here we focus demonstrating that Eq. (\ref{discrete}) is an exact PDE for the correlation function in the infinite volume limit, which provides asymptotically vanishing approximate solution in case of $\mathcal{L}=(1+\nabla^2)^2-\gamma$ (Swifth-Hohenberg/Brazowskij, or Phase-Field Crystal model), for which the energy density and the stability properties are calculable. Since true stochastic solutions display no long-range order, we take Eq. (\ref{discrete}) in the $L \to \infty$ (aperiodic) limit, which otherwise restores continuous rotation symmetry. Using $\sigma^2(q) \equiv \lim_{L \to \infty} [L/(2\pi)]^d \sigma^2_{\mathbf{q}} < \infty$, the inverse Fourier transform of Eq. (\ref{discrete}) reads:
\begin{equation}
\label{correlation}
[\hat{\mathcal{L}}+3(\bar{\varphi}^2+C_0)]^2\, C + (18\,\bar{\varphi}^2)\,C^2 + 6\,C^3 = \omega \enskip ,
\end{equation}
where $C(r) = \int d\mathbf{q}\, \{ \sigma^2(q)e^{\imath\,\mathbf{q}\cdot\mathbf{r}} \}$ is the correlation function of the coloured Gaussian random field, while $C_0=C(0)$. Furthermore, the condition $\sigma^2(0)=0$ indicates $\int_0^\infty d^dr\,C(r) =0$, while the lack of long-range order indicates $\lim_{r \to \infty} C(r) = 0$, thus yielding $\omega=0$. Any solution of Eq. (\ref{correlation}) satisfying the above conditions and $\sigma^2(k)\geq 0$ is a Gaussian solution of the Phase-Field Crystal (PFC) model. Henceforth we only consider the case $d=3$. The solution of Eq. (\ref{correlation}) then reads \cite{DavidSibley}:
\begin{equation}
\label{SHcorr}
C(r) = A^2 \,\gamma\,\frac{\sin(r)}{r} + O(\gamma^2) \enskip ,
\end{equation}
where $\gamma=\epsilon-3\,\bar{\varphi}^2 \geq$ is a small parameter (distance from the liquid stability line), while $A\,\sqrt{\gamma}$ is the characteristic amplitude of the random pattern. Since $\sigma(k) \propto A^2 \gamma \delta(k-1) \geq 0$, an approximating analytical solution exists in the leading order of $\gamma$, meaning that the realisations of a coloured Gaussian random field with correlation function $C(r)=A^2 \gamma \sin(r)/r$ are approximating analytical solutions to the Euler-Lagrange equation for some $A>0$ (to be determined later). Moreover, since $C(r)$ is smooth, $\phi(\mathbf{r})$ is also smooth in the realisations of the random field, and therefore Eq. (\ref{psik}) is valid \cite{gikhman1969introduction}. From the Fourier transform of Eq. (\ref{correlation}) it is clear that the solution terminates in a higher order of $\gamma$ [where $\sigma^2(q)\geq 0$ does not hold any more], but now we neglect this information and work further with the leading order solution to demonstrate the predictive power of the mathematical method. We also mention, that the only radial solution of Eq. (\ref{correlation}) is $C(r) \equiv 0$ for $\hat{\mathcal{L}}=-(1+\nabla^2)$ (Cahn-Hilliard model), which emerges from the fact that Eq. (\ref{correlation}) recovers the Euler-Lagrange equation of the Swift-Hohenberg model on the liquid linear stability curve in the parameter space, where radial solutions terminate \cite{Mccalla,Kulagin2008}.  

\subsubsection{Free energy density.}

The first step is to calculate the amplitude in Eq. (\ref{SHcorr}) for which the random configurations represent approximate analytical solutions to to EL equation in the leading order of $\gamma$. For this we need to calculate the expectation value and the variance of the energy density for the Gaussian phase. Using $\varphi=\bar{\varphi}+\phi$ and the Fourier representation of $\phi$ in $f_L:=F[\varphi]/L^3$ results in:
\begin{eqnarray}
\nonumber f_L(\Phi) = && f_0 + \frac{1}{2}\sum_{\mathbf{q}}\left[\mathcal{L}(q)+3\,\bar{\varphi}^2 \right](\phi^*_\mathbf{q}\phi_{\mathbf{q}}) \\
\label{energy}&&+ \bar{\varphi}\sum_{\mathbf{q}_1,\mathbf{q}_2,\mathbf{q}_3}\phi_{\mathbf{q}_1}\phi_{\mathbf{q}_2}\phi_{\mathbf{q}_3}\delta_{\mathbf{q}_1+\mathbf{q}_2+\mathbf{q}_3} \\
\nonumber && + \frac{1}{4}\sum_{\mathbf{q}_1,\mathbf{q}_2,\mathbf{q}_3,\mathbf{q}_4}\phi_{\mathbf{q}_1}\phi_{\mathbf{q}_2}\phi_{\mathbf{q}_3}\phi_{\mathbf{q}_4}\delta_{\mathbf{q}_1+\mathbf{q}_2+\mathbf{q}_3+\mathbf{q}_4} \enskip ,
\end{eqnarray} 
where $f_0=[(1-\epsilon)/2]\bar{\varphi}^2+\bar{\varphi}^4/4$ is the free energy density of the uniform phase of density $\bar{\varphi}$. The average free energy density can be calculated by taking the expectation value on Eq. (\ref{energy}), using Isserlis' Theorem to evaluate the terms $\langle \prod_{i=1}^m \phi_{\mathbf{q}_i}\rangle$, then taking the $L \to \infty$ limit, and using $\sigma^2(q)=1/(2\pi)^3 \int d^3x \, \{ C(r) \exp(-\imath\,\mathbf{q}\cdot\mathbf{r}) \}$, thus yielding:
\begin{equation}
\label{fenergy}
\bar{f} = f_0 + \frac{1}{2}\,\lim_{r \to 0} [(\hat{\mathcal{L}}+3 \, \bar{\varphi}^2)C(\mathbf{r}) ] + \frac{3}{4} \, C_0^2 \enskip .
\end{equation}
A similar calculation can be performed to determine the variance of the energy density $\sigma^2_{f} := \lim_{L \to \infty} \langle (f_L - \langle f_L\rangle)^2 \rangle$, 
yielding:
\begin{equation}
\label{variance}
\sigma_f^2 = \frac{1}{2}\,||(\hat{\mathcal{Q}}C)^2|| + (6\,\bar{\varphi})\,|| C^3 ||+\frac{3}{2}\,|| C^4 || \enskip ,
\end{equation}
\color{black}
where $\hat{\mathcal{Q}}=\hat{\mathcal{L}}+3(\bar{\varphi}^2+C_0)$, $||.|| = \lim_{L \to \infty} L^{-3}\int_{\Omega_L} d^3x\,\left\{.\right\}$ stands for the spatial average in the infinite volume limit. Since $\lim_{R \to \infty} R^{-3} \int_0^R \left[\sin(r)/r\right]^n r^2\,dr = 0$ for $n=2,3$ and $4$, $\sigma_f^2 \equiv 0$ for $C(r)\propto \sin(r)/r$. Since the free energy density of the approximate stochastic solutions is unique, they form a phase in the PFC model. Again, even if the solution of Eq. (\ref{correlation}) eventually terminates in a higher order of $\gamma$, the mathematical method is capable of predicting amorphous solid phases in continuum theories, which is a major achievement. Using now Eq. (\ref{SHcorr}) in Eq. (\ref{fenergy}) results in the Landau free energy $\bar{f} - f_0= (\gamma^2/2)\,A^2[ (3/2)A^2-1 ]$ with a maximum at $A=0$ (uniform) and two minima at $A=\pm 1/\sqrt{3}$ (approximate Gaussian), which indicates a spontaneous symmetry breaking (no first order transition) from the uniform phase to the Gaussian one. Since the cubic term of the free energy density is identically zero for the Gaussian measure, this result also accords with our expectations.\\

\noindent\textit{Comment: Independently from the particular probability measure, the convergence of the quadratic term in the expectation value of the energy density for infinite volume requires $\lim_{L \to \infty}\sum_{\mathbf{q}}\langle \phi^*_\mathbf{q} \phi_\mathbf{q}\rangle = \int d^3 s \,\sigma^2(q) = C(0) < \infty$. This is identical to the condition $\sigma^2(q)=\lim_{L \to \infty}\{[L/(2\,\pi)]^3\,\sigma^2_\mathbf{q} < \infty$, which sets the following scaling relation for the characteristic amplitude of the pattern: $\sigma^{(\lambda\,L)}_\mathbf{q}/\sigma^{(L)}_\mathbf{q} = (1/\lambda)^{3/2}$, where $\sigma_\mathbf{q}^{(L)}$ is the square root of $\sigma^2_\mathbf{q}$ to spatial period $L$. The equation means that the characteristic amplitude of the random pattern decreases with increasing system size, simply because the same energy content must be divided between more wave numbers. The connections between the wave numbers start to play a role in the higher order energy contributions. For $n$ even, the contribution of $\phi^{n}(\mathbf{r})$ to the expectation value of the free energy density reads: $g^{(L)}_n:=\langle \frac{1}{L^3}\int_\Omega d^dx \, \{ \phi^n(\mathbf{r})\} \rangle = \sum_{\mathbf{q}_1,\dots,\mathbf{q}_{n}} \langle \prod_{i=1}^{n} \phi_{\mathbf{q}_i} \rangle \delta_{\sum_{i=1}^{n}\mathbf{q}_i}$. If the Isserlis' theorem is applicable, the expected value reduces the number of closed polygons in the Fourier space contributing to the sum (see Section 4.1) by converting an integral of a power into a power of a finite integral in the infinite volume limit: $\lim_{L \to \infty} g_n^{(L)} \propto \lim_{L \to \infty} \left( \sum_{\mathbf{q}}\sigma^2_\mathbf{q} \right)^{n/2} = (C_0)^{n/2} < \infty$. In contrast, if all the polygons contributed to the expected value of the energy density, the scaling law for the energy contribution would be $g_n^{(\lambda\,L)}/g_n^{(L)} = \lambda^{3\left(\frac{n}{2}-1\right)}$. This converges only for $n=2$ (quadratic term), while diverges for any $n>2$ for $\lambda \to \infty$. Therefore, the convergence of the terms of the expected value of the free energy density in the infinite volume limit also imposes conditions on the possible solution(s) of Eq. (\ref{fullpsi}).}

\begin{figure}
	\centering\includegraphics[width=0.415\linewidth]{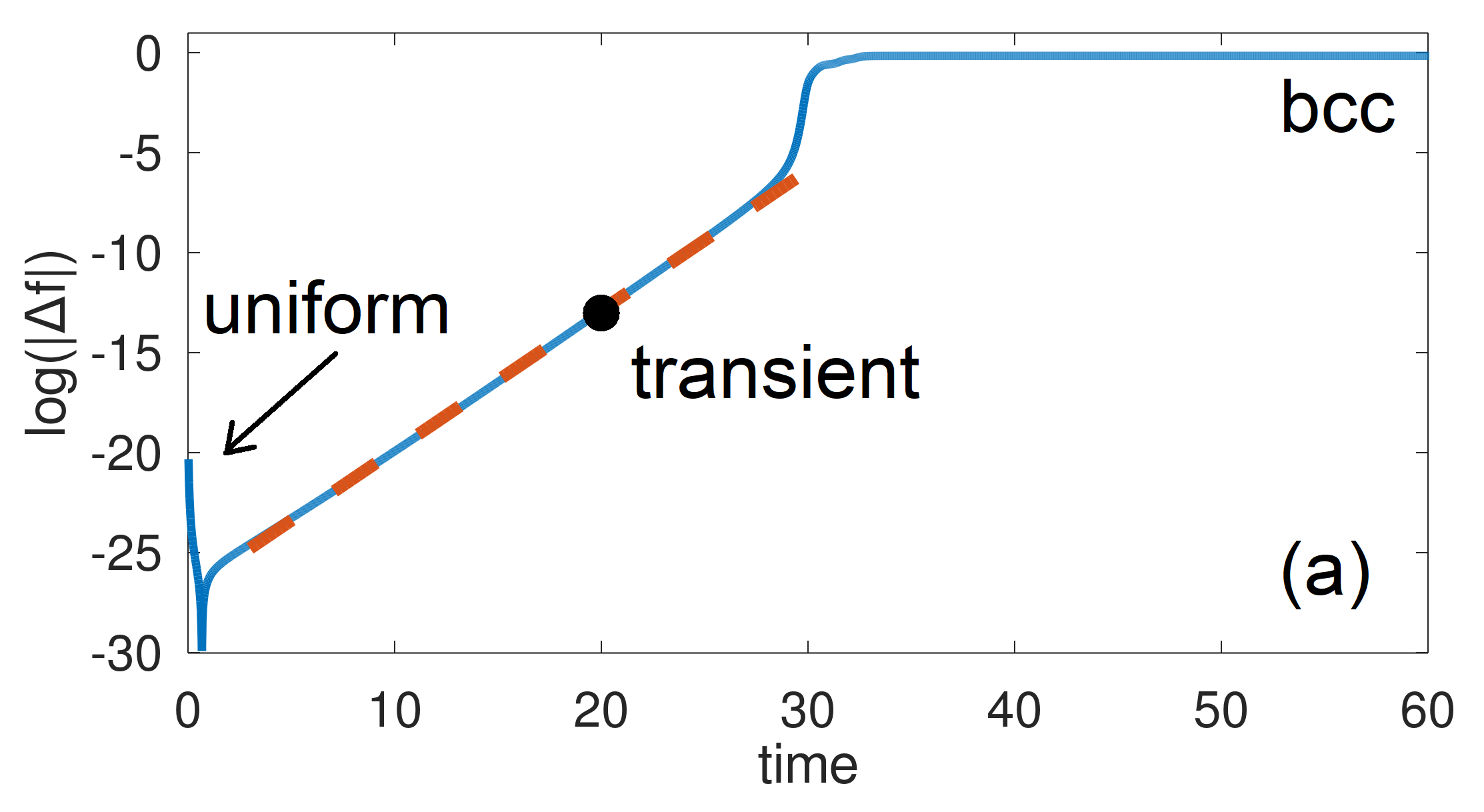}
	\includegraphics[width=0.274\linewidth]{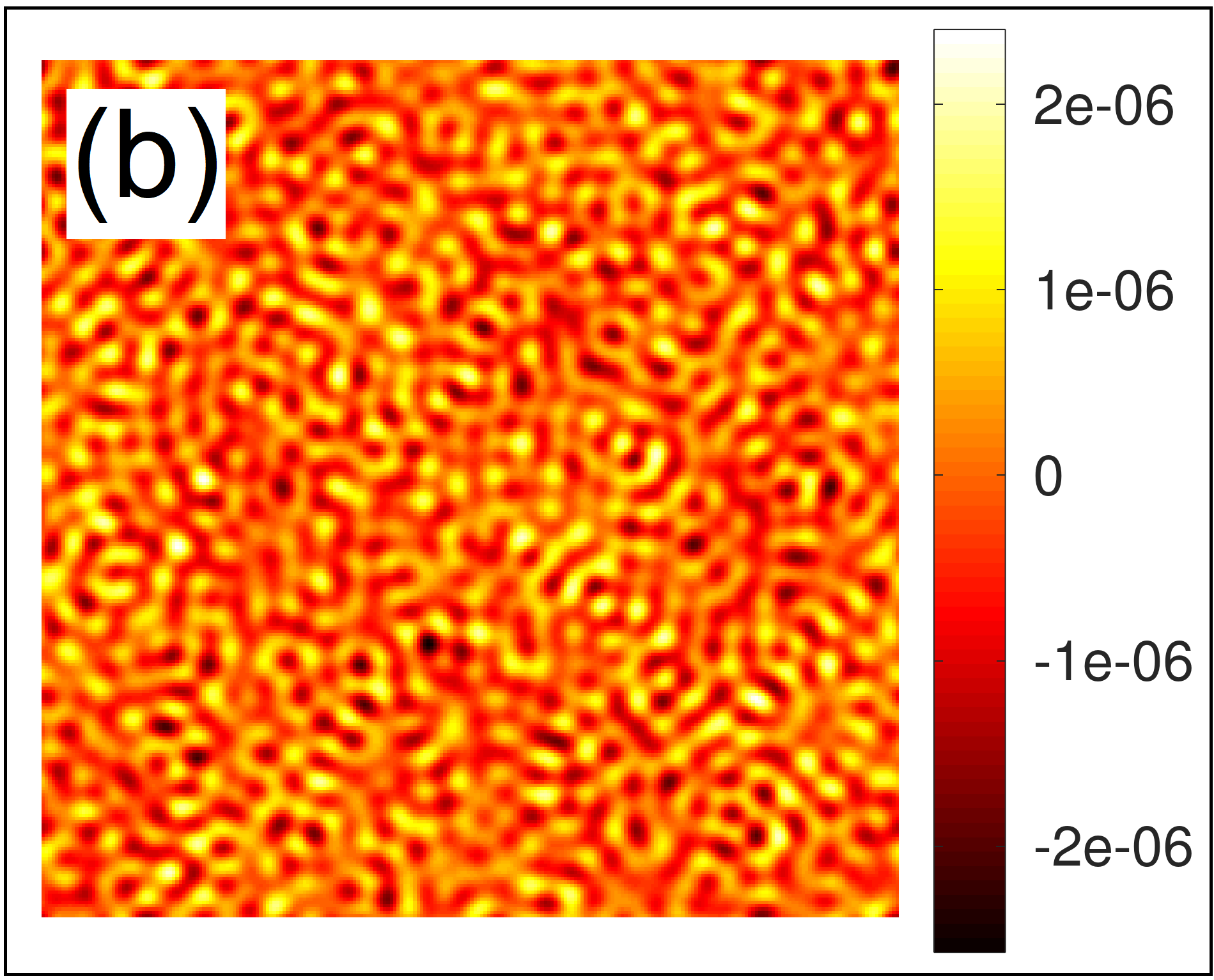}
	\includegraphics[width=0.25\linewidth]{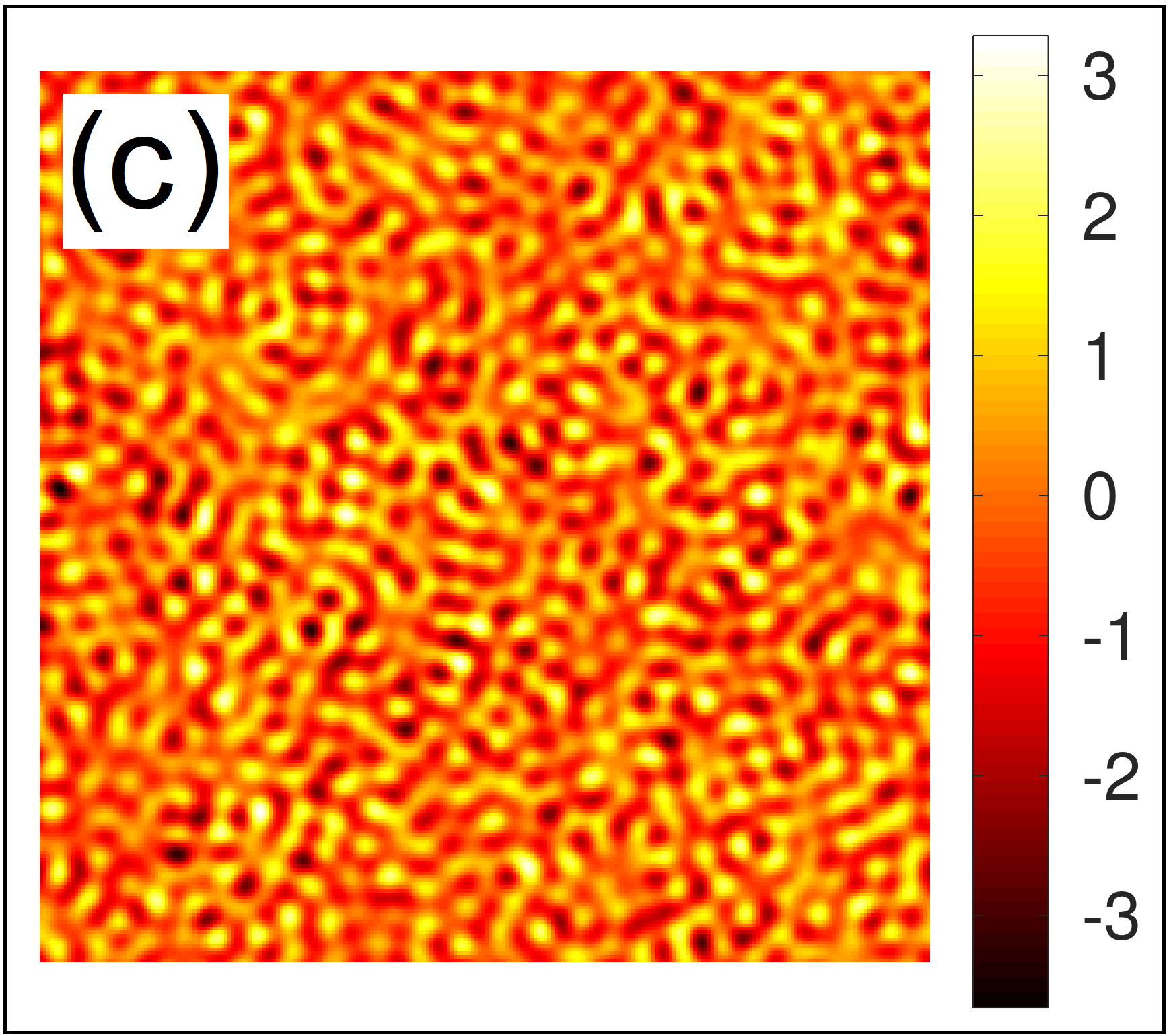}
	\caption{Gaussian transient state in the conserved Swift-Hohenberg (or Phase-Field Crystal) equation $\partial_t \phi=\nabla^2 \{[(1+\nabla^2)^2-\epsilon]\phi+(3\,\bar{\varphi})\phi^2+\phi^3\}$ at $\bar{\varphi}=-2.5\times10^{-2}$ and $\epsilon=2\times10^{-3}$ in a numerical simulation on a $256^3$ grid with grid spacing $h=2/3$ and time step $\Delta t=0.25$. The initial condition was $\phi(\mathbf{r},0)=10^{-3}\,\bar{\varphi}\,U[-1,1]$, where $U[-1,1]$ stands for a uniformly distributed random number on the interval $(-1,1)$. (a) Scaled dimensionless free energy density $\Delta f=10^7(f-f_0)$ vs scaled dimensionless time $t/10^5$. The dashed section of the curve indicates the \textit{exponential growth} of the amplitude of the transient pattern; (b) 2D cross section of the transient pattern at the point indicated on panel (a); (c) Numerical realisation of a coloured Gaussian random field with correlation function $C(r)\propto \sin(r)/r$. Note the structural similarity with panel (b).}
\end{figure}

\subsubsection{Stability.}

To determine the stability of the individual configurations in the approximate Gaussian phase, we start from the Taylor expansion of the energy density around a solution of the Euler-Lagrange equation, which reads:
\begin{equation}
\Delta f_L (\Phi,\delta\Phi) := f_L(\Phi+\delta\Phi) - f_L(\Phi) = \sum_{n=2}^4 \frac{1}{n!} \sum_{\mathbf{q}_1,\dots,\mathbf{q}_n}d^{(n)}_{\mathbf{q}_1,\dots,\mathbf{q}_n} \prod_{i=1}^n \delta \phi_{\mathbf{q}_i} \enskip ,
\end{equation}
where $d^{(n)}_{\mathbf{q}_1,\dots,\mathbf{q}_n} = [\partial^n f_L/(\partial\phi_{\mathbf{q}_1}\dots\partial\phi_{\mathbf{q}_n})]_{\Phi}$. Similarly to the free energy density calculations, we calculate the expectation value and the variance of the energy density difference $\Delta f_L (\Phi,\delta \Phi)$ for a fixed $\delta\phi(\mathbf{r}) := \sqrt{\gamma}\,\eta\,\psi(\mathbf{r})$ perturbation in the infinite volume limit, where the magnitude of the perturbation is measured in $\sqrt{\gamma}$ units (the characteristic amplitude of the approximate stochastic solutions to the EL), $O[\psi(\mathbf{r})]=1$, and $|\eta| \ll 1$. Since the free energy density of the Gaussian phase is unique, the calculations yield:
\begin{equation}
\label{dfbar} \overline{\Delta f} = \frac{1}{2}\,|| \delta\phi\hat{\mathcal{Q}}\delta\phi || + \bar{\varphi}\, || \delta\phi^3 || + \frac{1}{4}\,|| \delta\phi^4 ||
\end{equation}
and
\begin{eqnarray}
\nonumber \sigma_{\Delta f}^2 = && (3\,\bar{\varphi})^2\,|| \delta\phi^2 (C \circledast \delta\phi^2) || + (6\,\bar{\varphi})\,|| \delta\phi^3(C \circledast \delta\phi^2) || \\
\label{dfsigma} && + \frac{9}{2}\, || \delta\phi^2(C^2 \circledast \delta\phi^2) || + || \delta\phi^3(C \circledast \delta\phi^3) || \enskip , \quad \quad
\end{eqnarray}
where $(f \circledast g)(\mathbf{r}) = L^{-3}\int_{\Omega_L} d^3x'\{f(\mathbf{r}')g(\mathbf{r}-\mathbf{r}')\}$. Assuming $0 < || \psi^2(C \circledast \psi^2)|| < \infty$, the leading order of Eq. (\ref{dfsigma}) reads $\sigma_{\Delta f}^2 \propto \bar{\varphi}^2\,\gamma^3 \eta^4$. The Gaussian phase is stable if (but not only if) $\overline{\Delta f}>0$ and $\sigma_{\Delta f}^2=0$ for any non-trivial perturbation. Since $\mathcal{Q}(k)=(1-k^2)^2$ in the leading order of $\gamma$ for Eq. (\ref{SHcorr}), $\overline{\Delta f} \propto \gamma\,\eta^2 > 0$ in the leading order for any perturbation containing at least one $k \neq 1$ Fourier mode. The corresponding variance of the relative energy response $||\Delta f||:=\Delta f/\overline{\Delta f}$ is $\sigma^2_{||\Delta f||}=\sigma_{\Delta f}^2/\overline{\Delta f}^2 \propto \bar{\varphi}^2\,\gamma$. Assuming normal distribution for $\Delta f$, $\mathbb{P}(\Delta f \leq 0)=\frac{1}{2}\textrm{erf}[1/(\sigma_{|| \Delta f||} \sqrt{2})]$, and therefore there exists a finite $\gamma_0>0$ for any $\delta \phi$ so that the Gaussian phase is practically stable for $0<\gamma < \gamma_0$ against perturbations containing at least one $|\mathbf{k}| \neq 1$ Fourier mode. For perturbations consisting of only $|\mathbf{k}|=1$ modes, however, the quadratic term $\frac{1}{2}\,|| \delta\phi\,\hat{\mathcal{Q}}\,\delta\phi || = \frac{1}{2}\sum_{\mathbf{k}}\mathcal{Q}(k)|\delta\phi_{\mathbf{k}}|^2$ cancels in Eq. (\ref{dfbar}), which directly emerges from $\mathcal{Q}(1)=0$. For perturbations providing $|| \psi^3 || \neq 0$, $\overline{\Delta f} \propto \bar{\varphi}\,\gamma^{3/2}\eta^3$, which can be negative. It is easy to see that $|| \psi^3 ||>0$ for the bcc and 2D hexagonal structures, and therefore $\overline{\Delta f}<0$ for $\bar{\varphi}<0$ and $\eta>0$, meaning that the approximating Gaussian phase is unstable against perturbations displaying the structure of the stable phase of the system. Consequently, our analytical results predict an unstable Gaussian phase in the PFC model in the leading order of $\gamma=\epsilon-3\,\bar{\varphi}^2 \geq 0$, thus confirming the absence of this phase and the collapse of the Gaussian pattern into the stable bcc phase in time-dependent numerical simulations (see Figure 3).

\section{Conclusions}

Motivated by the need for a theoretical understanding of the fundamental nature of spatially non-uniform states of matter with no long-range order, a mathematical framework has been developed to investigate the existence of stochastic solutions to time-independent partial different equations (PDE's). The central idea of the method is the measure theoretical re-formulation of PDE's, which relies on the limit case of the concentration of measure phenomenon, when functions of infinitely many random variables become constant. A stochastic solution to a PDE is defined as a set of infinitely many solutions selected by a non-singular probability measure from a measurable function space. If there exist an associated energy density so that the PDE provides its stationary points, and the energy is unique over the individual configurations in the stochastic solution, the stochastic solution represents a spatially non-uniform physical phase with no long-range order (called amorphous solid phase in general), since the energy of the phase is the analytical function of the macroscopic state variables. The method was implemented continuum mean-field theories of first order phase transitions with quartic non-linearity, where it has been found that the only fully symmetry preserving stationary point of the energy is the constant solution. In addition to this exact result, the practicability and predictive power of the concept have also been demonstrated by finding an approximating stochastic solution in the leading order of a small parameter in the Phase-Field Crystal model, for which the energy density and the stability properties were also determined. It has been found that our method is capable of predicting amorphous solid phases, and is applicable to determine the free energy density and stability properties of these. Consequently, if we accept that regular analytical solutions (homogeneous, crystal lattice symmetric, etc) of PDE's represent physical phases, we must accept that stochastic solutions also do. Regarding the existence of these solutions, our results suggest that symmetry breaking must be present in stochastic solutions on the level of the probability measure (rather than on the level of the individual ``random'' solutions themselves, which seem to have no symmetry at all). With this idea, it would be possible to find and classify amorphous solid phases in the spirit of symmetry breaking. We note here that the idea of the presence of crystal symmetries in amorphous solids in some form is not completely new. A simple crystal randomisation process was developed and successfully applied to recover the experimental radial distribution function of metallic glasses \cite{qcalloys,VINCZE1980499}, and the presence of short-range order in amorphous solids is also supported by recent computer simulations \cite{PhysRevLett.107.175702}.

The next step of the research is to find the stable stochastic solution in the Phase-Field Crystal model shown in Figure 1(b). We believe that this is possible via a comprehensive statistical analysis of large (in the order of 10.000) number of convergent configurations in numerical simulations, which indeed necessitates substantial HPC capacity. In theory, a measure can be re-constructed from its moments under certain circumstances \cite{moment}. However, we do not expect the need for a complete computational reconstruction process from scratch, since it might be enough to validate our symmetry breaking arguments and assumptions made to find the essential components of the measure. Such a study is already in progress. The method will also be applied for more quantitative mean-field theories (such as the Classical Density Functional Theory of soft matter and the Fundamental Measure Theory), where conditions for the existence of the amorphous solid phase will be given in terms of physical quantities. Since we search for disordered analytical solutions of Euler-Lagrange equations directly (instead of minimising only a parametrised free energy density functional, which might miss the actual stationary points), this step is expected to result in a major progress in glass transition theory. In addition, developing the analogue of the Kramers/Fokker-Planck equation governing the time evolution of parameters of probability measures on continuum dynamics' is also vital. Finally we note that the general mathematical idea is not restricted to continuum models, since the only requirement on the level of the mathematical model is an (at least) countable infinitely many dimensional representation of the physical state of the system. Consequently, our results are expected to have important implications in all fields of Condensed Matter research where non-uniform states with no long range order emerge.

\section*{Acknowledgements}

The author wishes to thank David N. Sibley, Wael Bahsoun, Tapio Ala-Nissil\"a, Andrew J. Archer, L\'aszl\'o Gr\'an\'asy and Frigyes Podmanizcky for their valuable comments and support. 

\section*{References}

\bibliographystyle{unsrt}
\bibliography{./papers}

\end{document}